\documentclass[twocolumn,amsmath,amssymb]{revtex4}

\usepackage{graphicx}
\usepackage{dcolumn}
\usepackage{bm}

\begin{document}
\title{Electronic parameters for the hole transfer in DNA duplex oligomers.} 
\author{Julia A. Berashevich and Tapash Chakraborty$^{\ast}$}
\affiliation{Dr. J. A. Berashevich, Prof. T. Chakraborty \\
Department of Physics and Astronomy\\
The University of Manitoba \\
Allen Building, R3T 2N2, Winnipeg (Canada) \\
E-mail: tapash@physics.umanitoba.ca}

\begin{abstract}
We report on our calculations of the inner-sphere reorganization 
energy and the interaction of the $\pi$ orbitals within DNA oligomers. 
The exponential decrease of the electronic coupling between the 
highest and second highest occupied base orbitals of the intrastrand 
nucleobases in the (A-T)$_n$ and (G-C)$_n$ oligomers have been found 
with an increase of the sequence number $n$ in the DNA structure.  
We conclude that for realistic estimation of the electronic coupling 
values between the nucleobases within the DNA molecule, 
a DNA chain containing at least four base pairs is required.
We estimate the geometry relaxation of the base pairs
within the (A-T)$_n$ and (G-C)$_n$ oligomers ($n=1-6$) due to their oxidation. 
The decrease of the inner-sphere reorganization energy with elongation of the 
oligomer structure participating in the oxidation process have been observed.
The maximum degree of geometry relaxation of the nucleobase structures 
and correspondingly the higher charge density 
in the oxidized state are found to be located close to the oligomer center. 
\end{abstract}

\maketitle
\section{Introduction}
Discovery of charge migration in DNA molecules has 
opened new avenues to investigate various possibilities ranging 
from its role in the DNA oxidative damage and repair \cite{oxidative} 
to application of DNA in nanoelectronic device developments 
\cite{kawai}. In fact, DNA-based molecular electronic devices 
are expected to operate within the picoseconds range \cite{wan,kel} 
that can exceed the potential of the present solid state devices. 
Quite expectedly, the DNA molecule has become a subject of intense 
research activities both theoretically 
\cite{ratner,berash,siguyam,voit,troisi,roche,chakra_group} 
and experimentally \cite{gieN,lewis1,Schuster,hess,neil}. 

From all these studies of charge migration in the DNA 
molecule reported as yet, it is clear that there are two mechanisms 
for transfer of charge depending on the DNA structure and transfer parameters: 
a superexchange charge transfer and the incoherent hopping \cite{gieN}. The 
charge migration leads to the geometry changes in the nucleotides and the 
surrounding environment, which significantly contribute to the charge 
migration process. Due to the interaction of the $\pi$ orbitals of the 
nearest neighbor duplexes and insignificant IP difference between them, 
hole can be distributed over several sites in the (A-T)$_n$ and (G-C)$_n$ 
oligomers. This significantly changes the magnitudes of the geometry 
relaxation of the nucleobases -- inner-sphere component and environment 
contribution -- outer-sphere component. However, the investigation of 
the transfer parameters, such as orbital overlapping \cite{troisi,senthil,voit} 
and activation energy for charge migration i.e. the IP and the reorganization 
energy \cite{siguyam,troisi,senthil,voit,tav,ton,olof}, have been performed mostly for the nucleobases or/and base pairs.

The main purpose of our work is to estimate the 
electronic coupling between the two nearest nucleobases, their
charge distribution and inner-sphere reorganization energy, 
when they are placed within the (A-T)$_n$ and (G-C)$_n$ oligomer duplexes.  
All these computations have been performed 
using accurate quantum-chemical methods. 

\section{Method of computation}
The relatively small reaction free energy in the DNA molecule makes the DNA
hole transfer mechanism qualitatively different from that in most proteins
\cite{hush} . The electron transfer in the DNA molecule was found to be 
strongly dependent on the details of the donor and acceptor energies and 
deviation of their geometries \cite{neil}.

The charge transfer in a DNA molecule occurs due to the 
overlapping between the $\pi$-electrons of the carbon and 
the nitrogen atoms that forms the $\pi-\pi$ orbitals between 
the parallel nucleobases. Charge migration in the molecular 
systems with weakly interacting donors and acceptors, such as 
between the base pairs in the DNA molecule, is described by 
the standard high-temperature nonadiabatic electron-transfer 
rate 
\begin{equation}
k=\frac{2\pi}{\hbar}|H_{DA}|^{2} (FC),
\label{eq:rate}
\end{equation}
where $H_{DA}$ is the electronic donor-acceptor matrix element, 
and FC is the Franck-Condon factor. 

The electronic donor-acceptor matrix element $H_{DA}$ is defined by the 
coupling of the orbitals of the donor and the acceptor and depends 
on the structure of the DNA molecule. For the (A-T)$_n$ and (G-C)$_n$ 
oligomers the simple expressions for the deviation of the electronic 
coupling on the sequence number $n$ have been generated \cite{starikow}. 
According to these expressions, the value of the electronic coupling 
decreases with elongation of the oligomers \cite{starikow}. In Sect.~III A, 
we simulate the electronic coupling of the nucleobases within the (A-T)$_n$ 
and (G-C)$_n$ oligomers using the quantum chemistry methods with the 
Jaguar 6.5 program \cite{jaguar}. According to the Koopmans' theorem, the 
electronic coupling can be estimated as half of the adiabatic state 
splitting between the HOMO and the HOMO-1 of the closed shell neutral system, 
determined in a Hartree-Fock self-consistent field. Therefore, the 
RHF/6-31$^{+}$G$^{\ast}$  have been applied for the electronic coupling 
calculations. The 6-31$^{+}$G$^{\ast}$ basis set is appropriate for our 
purposes. Previous investigations indicated that any further extension 
has little influence on the electron coupling \cite{voit}. The geometries 
of the separated DNA base pairs have been optimized with the 
RHF/6-31$^{+}$G$^{\ast}$ bases and in the following, the optimized 
geometries of the base pairs have been stacked with a twist angle 
36$^{\circ}$ and a distance of 3.38 \r{A}. The stacking of the preliminary 
optimized geometries allows us to consider the same nucleobases to be 
`in resonance' \cite{voit} within the structures of the (A-T)$_n$ and 
(G-C)$_n$ oligomers. 

The FC factor deals with the influence of the vibronic interaction 
on the charge propagation and can be expressed as
\begin{equation}
\mathrm{FC}=(4\lambda k_BT)^{-\frac12}\exp\left(-\frac{(\Delta G+
\lambda)^2}{4\lambda k_BT}\right),
\label{eq:FC}
\end{equation}
where $\Delta G$ is the free energy of the reaction, and $\lambda=
\lambda_s+\lambda_i$ is the reorganization energy. The interaction 
of the molecule with the solvent environment is included in the 
outer-sphere reorganization energy  $\lambda_s$, while the relaxation 
of the acceptor, the donor and the molecular bridge geometries are 
included in the inner-sphere reorganization energy $\lambda_i$. 

\begin{figure}
\includegraphics[scale=0.9]{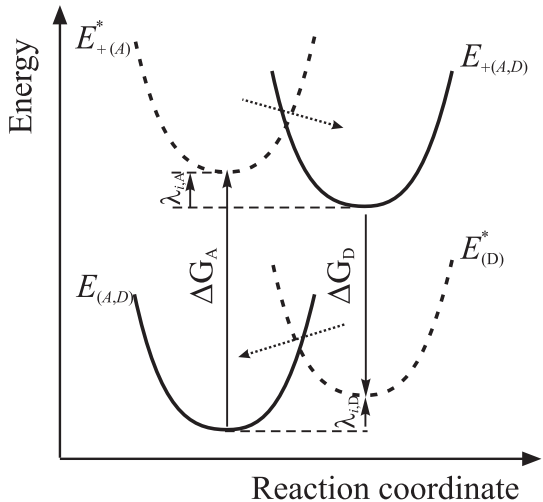}
\caption{Schematic diagram for the Marcus-Hush scheme of the 
inner-sphere reorganization energy: (a) the reorganization energy 
$\lambda_{i,D}$ for adding one electron to the positive ion and (b) 
the reorganization energy $\lambda_{i,A}$ for removing one electron 
from the neutral geometry, 
where $D$ and $A$ are the hole donor and the hole acceptor.}
\label{fig:fig1}
\end{figure}

The inner-sphere reorganization energy accounts for the low-frequency 
inter-molecular modes and can be estimated within the quantum chemical 
approach as 
\cite{mark2}
\begin{equation}
\lambda_i=\lambda_{i,A}+\lambda_{i,D}=(E^{\ast}_{+(A)}-E_{+(A)})+
(E^{\ast}_{(D)}-E_{(D)}),
\label{eq:five}
\end{equation}
where $E$ is the energy of the neutral state in a neutral geometry, 
$E^{\ast}$ is the energy of the neutral state in an ionic geometry, 
$E_+$ is the energy of the ionic state in an ionic geometry, and 
$E^{\ast}_+$ is the energy of the ionic state in a neutral geometry. 
The reorganization energy $\lambda_{i,A}$ is the energy to remove 
an electron from the hole acceptor $A$, while the reorganization 
energy $\lambda_{i,D}$ is the energy to add an electron to the hole 
donor $D^+$. The scheme for calculation of the reorganization energy 
is presented in Figure~\ref{fig:fig1}. Clearly, the vibronic interactions 
stabilize the geometry of the donor and the acceptor from a non-equilibrium 
state ($E^{\ast}_{+(A)}, E^{\ast}_{(D)}$) to the equilibrium state 
($E_{+(A)},E_{(D)}$). The vertical ionization potential is determined 
as $\mathrm{vIP}=(E^{\ast}_{+}-E)$ and differ from the adiabatic
$\mathrm{IP}=(E_{+}-E)$ by the inner-sphere reorganization energy.

The inner-sphere reorganization energy has been evaluated within 
the unrestricted Becke3P86/6-311$^{+}$G$^{\ast}$ approximation of 
the DFT method. The DFT theory was found to be reasonable for this 
purpose based on a comparison of the results of Ref.~\cite{olof}. 
These results show that the DFT theory predicts the magnitude of 
the inner-sphere reorganization energy with a minimum error when 
compared to the experimental data \cite{hush,sevi,orlov}. Furthermore, 
we have also tested the application of the HF method and the DFT theory 
for the vertical ionization potential (vIP) calculations and have 
found significant qualitative and quantitative disagreement of the 
HF with the experimental data \cite{hush}, while the Becke3P86 
approximation is appropriate for this purpose. 

\section{Results and discussion} 
\subsection{High occupied orbital distribution}
At first we consider the system of two stacked duplexes. The results 
for the highest occupied base orbital (HOBO) are presented in 
Table~\ref{tab:table2}. In the case when the pyrimidine/pyrimidine 
and purine/purine bases are stacked in one strands, the HOBOs of the 
adenine and guanine bases have lowest energy in comparison to the 
pyrimidine/purine configurations. For the (A-T)$_2$ and (G-C)$_2$ 
oligomers the HOBOs are delocalized over the two intrastrand nucleobases, 
and therefore, it produces a significant coupling between the $\pi$ 
orbitals of the stacked pyrimidine/pyrimidine and purine/purine bases. 
For oligomers where the pyrimidine and the purine bases are stacked 
in the same strand (A-T/T-A, G-C/C-G, A-T/C-C) or in the mixed structures 
(A-T/G-C and G-C/A-T), for some cases the $\pi$ orbitals are delocalized, 
but electronic coupling is weak. For others the $\pi$ orbitals are 
localized mostly on one nucleobase, and we can consider the weak 
intrastrand and interstrand coupling between the nucleobases as well. 
The low interstrand coupling for the cases A-T/T-A and G-C/C-G has been 
observed experimentally \cite{neil1}.

\begin{table*}
\caption{\label{tab:table2} The HOBO within the system of two 
stacked base pairs estimated with RHF/6-31$^{+}$G$^{\ast}$//RHF/6-31$^{+}$G$^{\ast}$,
in the case when the orbital is localized (l) on a single 
nucleobase, or delocalized (d). All values are in eV.}
\begin{ruledtabular}
\begin{tabular}{cccccccc}
 & A A  & A T  & G G  & G C  & A G  & G A  & A C\\
 &$|$ $|$ &$|$ $|$ &$|$ $|$ &$|$ $|$ &$|$ $|$ &$|$ $|$ &$|$ $|$\\ 
 & T T  & T A  & C C  & C G  & T C  & C T  & T G \\ 
\hline
adenine & 8.10(d),8.38(d)& 8.17(d),8.33(d) &  - & -               
& 7.96(l) & 8.24(l) & 8.32(l) \\
thymine & 9.24(d),9.71(d)& 9.01(l),9.35(l) &  - & -               
& 9.38(d) & 9.48(d) & 8.69(l) \\
guanine & - & - & 7.17(d),7.48(d) & 7.46(l),7.77(l) 
& 7.39(l) & 7.34(l) & 7.44(l) \\
cytosine& - & - & 9.69(d),10.08(d) & 9.33(l),9.36(l) 
& 9.71(d) & 9.81(d) & 9.65(l) \\
\end{tabular}
\end{ruledtabular}
\end{table*}

\begin{figure}
\includegraphics[scale=1.2]{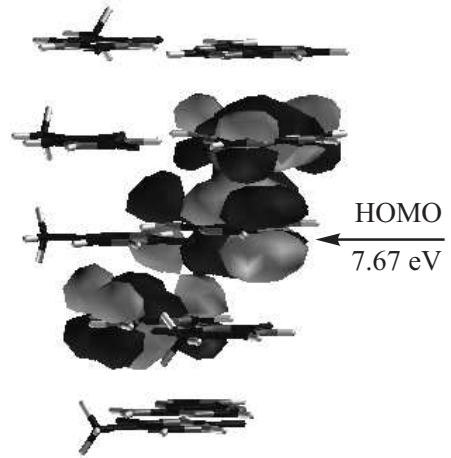}
\caption{The electronic density of the HOMO that resides on the 
central adenine (as indicated) in the (A-T)$_5$ DNA sequence 
(RHF/6-31$^{+}$G$^{\ast}$//RHF/6-31$^{+}$G$^{\ast}$).}
\label{fig:fig4} 
\end{figure}

In the (A-T)$_n$ and (G-C)$_n$ DNA oligomers, the 2$n$
$\pi$ orbitals are delocalized over the 
neighboring intrastrand nucleobases.   
We have found that the HOMO corresponds to the central nucleobases 
in the (A-T)$_n$ and (G-C)$_n$ oligomers structures. 
Particularly, the HOBOs of the adenine and guanine primarily
belongs to the nucleobase in the center of the oligomer, 
while HOBO-1 belongs to the nearest neighboring nucleobase 
to that in the oligomer center.
Similarly, the HOMO resides primarily on the central guanine 
have been observed for the 5'-(G-C)$_n$ structures \cite{lebreton}. 
This effect is related to the electrostatic potential distribution 
over the (A-T)$_n$ and (G-C)$_n$ oligomers in the vacuum.
We have computed the electrostatic potential distribution in the 
(A-T)$_n$ and (G-C)$_n$ oligomers with the the APBS program performing the 
nonlinear Poisson-Boltzmann solver \cite{baker}. 
The RESP procedure \cite{koll} has been applied to determine an  
atomic partial charge of the A-T and G-C base pairs. 
The atomic partial charges of the base pairs have been 
the same for each base pair in the oligomer structures. 
We have found that for the (A-T)$_n$ structure, the 
the electrostatic potential on the central adenine and thymine
is more negative than on the nucleobases of the oligomer sides.
For the (G-C)$_n$ structure, the electrostatic potential on 
central guanine is more negative than on the sides, while for the 
cytosine the opposite effect takes place. Therefore, the stronger 
localization of the HOMO density on the central nucleobase is observed 
for the (A-T)$_n$ oligomer, where the HOMO is mostly delocalized over 
three neighboring intrastrand adenines, than for the (G-C)$_n$ oligomer, 
where the HOMO is delocalized over four neighboring guanines. The HOMO 
electronic density for the (A-T)$_5$ DNA sequence are presented in 
Figure ~\ref{fig:fig4}. It has been found, that for the $n$=3 the 
population of the HOMO is much larger for the central nucleobases than 
for the sides.  For the $n>3$ the population of the HOMO 
on the central nucleobases decreases due to the HOMO delocalization. 

\begin{figure}
\includegraphics[scale=0.9]{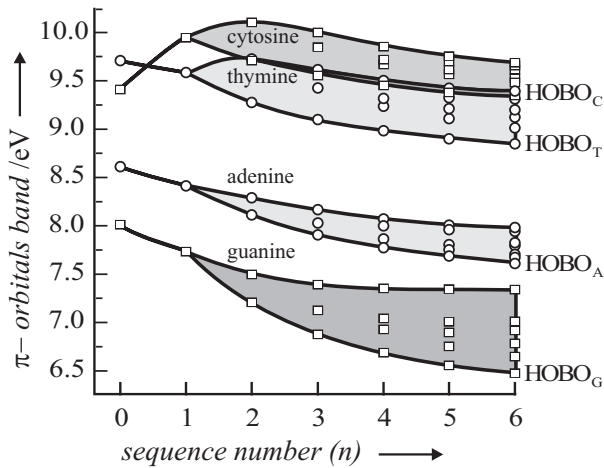}
\caption{The $\pi$ orbitals of the nucleobases within the (A-T)$_n$ and 
the (G-C)$_n$ oligomers, where $n$ is number of the DNA base pair 
(RHF/6-31$^{+}$G$^{\ast}$// RHF/6-31$^{+}$G$^{\ast}$).}
\label{fig:fig5} 
\end{figure}

\begin{figure}
\includegraphics[scale=0.9]{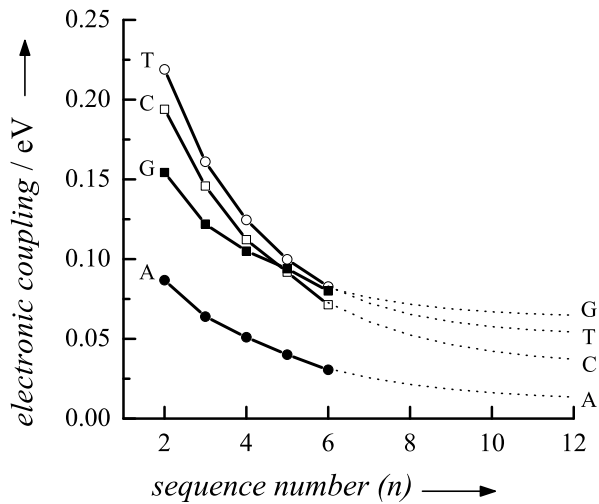}
\caption{The electronic coupling between the HOBO and HOBO-1 for 
the nucleobases in the (A-T)$_n$ and the (G-C)$_n$ duplex oligomers 
(RHF/6-31$^{+}$G$^{\ast}$// RHF/6-31$^{+}$G$^{\ast}$). 
The dotted lines correspond to the extrapolated results.}
\label{fig:fig6} 
\end{figure}

The delocalization of the orbital electron density 
over the oligomer structure produces a decrease 
of the HOBO energies with elongation of the DNA chain. 
The dependence of the orbital energies of the $n$ HOBOs, 
which are $\pi$ orbitals of the nucleobases, in the 
(A-T)$_n$ and (G-C)$_n$ oligomers on the sequence number 
$n$ are presented in Figure ~\ref{fig:fig5}. As we see from 
the results, the splitting of the HOBO and HOBO-1 decreases with 
elongation of the duplex oligomer structures. We conclude that 
the electronic coupling between the HOBO and HOBO-1
belonging to the nearest intrastrand nucleobases decreases as well
due to the spreading of the electron density of the molecular $\pi$ orbitals 
over the larger sequence number. Therefore, the maximum value of 
the electronic coupling is observed for the structures of the two 
stacked base pairs, where the HOBO can spread only over two nucleobases. 
The electronic coupling between the two HOBO and HOBO-1 for the different 
nucleobases within the (A-T)$_n$ and (G-C)$_n$ oligomers are presented in 
Figure ~\ref{fig:fig6}. As we mentioned above, the $\pi$ orbitals 
are primarily delocalized over 3-4 intrastrand nucleobases. Therefore, 
a fast decrease of the electron coupling for $n\le$4 is observed when 
the $\pi$ orbitals have the potential to spread (see Figure~\ref{fig:fig4}). 
For $n>4$ the electronic coupling decreases slowly. 
For $n$=6, the electronic coupling magnitude is less than half of that for $n$=2. 
According to the performed extrapolation procedure in Figure ~\ref{fig:fig6}, for 
$n\ge$8 the coupling is practically independent of the sequence number.

The observed decrease of electronic coupling in the (A-T)$_n$ and (G-C)$_n$ 
oligomers is in good agreement with the approximation in Ref.~\cite{starikow},
where for $n$=4 the electronic coupling is half of that for $n$=2.
Moreover, it can explain the disagreement of the earlier theoretical 
results \cite{voit,senthil} and experiments, where the electronic coupling 
is usually much smaller than 0.1 eV \cite{lewis1}. Based on these 
results we conclude that the electronic coupling calculated for the system 
of two stacked base pairs leads to the overestimation of the transfer 
integral for the charge transfer simulation in the DNA oligomers. It seems 
to us that more accurate value of the electron coupling can be 
determined from the DNA chain with length of at least four base pairs.

\subsection{Inner-sphere reorganization energy}
The orbital overlapping and the distribution of the HOMOs 
over the oligomer structure directly influence
the charge distribution in the DNA molecule 
and the inner-sphere reorganization energy, respectively.

We have calculated the
inner-sphere reorganization energy and the vIP for the separated 
nucleobases and their pairs. The results are compared with the 
experimental data in Table~\ref{tab:table1}.  
From Table~\ref{tab:table1} it is clear that our calculation and 
the experimental results for the nucleobases are in good agreement 
except for cytosine. Unfortunately, the experimental values were 
defined as the difference between the vertical ionization 
potential and the adiabatic ionization potential, 
with values extracted from different sources \cite{hush,sevi,orlov}. 
As a result, in general the final values can be inaccurate due to 
the use of different experimental techniques and agreement of the 
theoretical data with the experiment is not completely reliable. 
For the nucleobases, the large geometry relaxation between the 
neutral and oxidized states, particularly the change in C-C, C-N 
and C-O bond lengths, is the cause of the large magnitude of the 
inner-sphere reorganization energy. In the case of the pair 
formation, the inner-sphere reorganization energy is defined as 
the geometry relaxation of both nucleobases of the pair and is decreased by the 
flexibility of the hydrogen bonds in the neutral and ionic geometries. 
The unit charge, spread between the two nucleobases in the pair 
instead of one, decreases the geometry relaxation of each nucleobase as well. 
For the A-T pair the inner-sphere reorganization energy is found 
to be one-third of the sum of the reorganization energy of the 
separated adenine and thymine. That is a result of insignificant 
geometry relaxation of the nucleobases itself during the oxidation process 
because of the high flexibility 
of the two hydrogen bonds (opening translation \cite{starikow}). There 
are three hydrogen bonds between the nucleobases in a G-C pair, 
which restrict the translation and rotation flexibility of the 
nucleobases. This causes the not so significant decrease of the 
inner-sphere reorganization energy for the G-C pair 
compared to that of the A-T pair. 

\begin{table}
\caption{\label{tab:table1} The inner-sphere reorganization energy 
and the values of vIP for the nucleobases and pairs estimated with 
UB3P86/6-311$^{+}$G$^{\ast}$. All values are in eV.}
\begin{ruledtabular}
\begin{tabular}{ccccccc}
 &$\mathrm{vIP}$ &$\mathrm{vIP}$\footnotemark[1] &$\lambda_{i,A}$
 &$\lambda_{i,D}$ &$\lambda_{i}$ &$\lambda_{i,A}$\footnotemark[2]\\
\hline
adenine& 8.8740 & 8.44 & 0.1999 & 0.2020 & 0.4019 & 0.18 \\
thymine& 9.3797 & 9.14 & 0.2432 & 0.2703 & 0.5135 & 0.27 \\
guanine& 8.4626 & 8.24 & 0.4344 & 0.4392 & 0.8737 & 0.47 \\
cytosine& 9.3049 & 8.94 & 0.1107 & 0.1120 & 0.2227 & 0.26 \\
A-T& 8.4381 & - & 0.1492 & 0.2184 & 0.3676 & -  \\
G-C& 7.8326 & - & 0.3526 & 0.3679 & 0.7205 & - \\
\end{tabular}
\end{ruledtabular}
\footnotetext[1]{Experimental data \cite{hush}}
\footnotetext[2]{Experimental data \cite{hush,sevi,orlov}}
\end{table}

\begin{figure}
\includegraphics[scale=0.9]{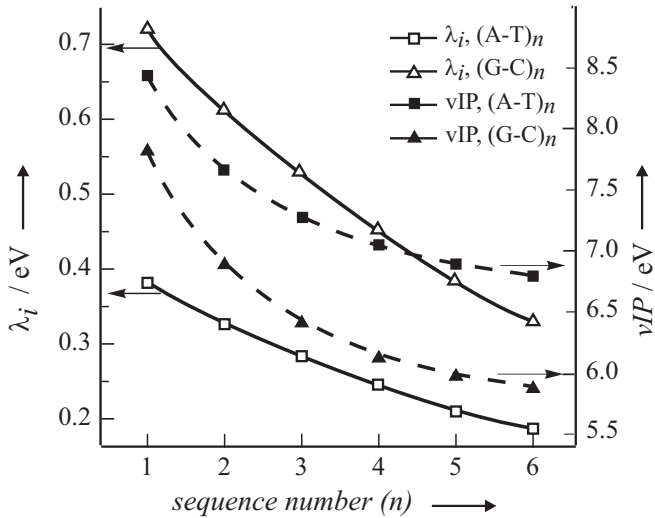}
\caption{The inner-sphere reorganization energy $\lambda_{i}$ and 
the vIP values versus the number of pairs in a DNA duplex oligomers 
(A-T)$_n$ and (G-C)$_n$ performed with UB3P86/6-311$^{+}$G$^{\ast}$} 
\label{fig:fig2}
\end{figure}

Further, the hydrogen bonds are the channels for charge transfer 
between the nucleobases. In the oxidized state the hydrogen bonds 
participate in the charge transfer between the nucleobases to bring 
the pairs from the nonequilibrum state, where the charge is localized 
only on the nucleobase with a lower IP, to the equilibrium state, 
where the charge is spread over the base pair \cite{guerra}. 
We consider the A-T and G-C base pairs as a single state 
for the following calculations of $\lambda_i$.
The stacking of the base pairs into the (G-C)$_n$ and (A-T)$_n$ 
oligomers leads to a decrease of the inner-sphere reorganization energy 
$\lambda_{i}$ and a decrease of the vIP as well. The results are 
presented in Figure ~\ref{fig:fig2}, where the decrease of $\lambda_i$ 
is seen to occur due to the contribution 
of the rotation and translation of the base pairs relative to each 
other and to the spreading of the charge between the pairs. 
According to our data, with elongation of the (A-T)$_n$ and (G-C)$_n$ 
oligomers the twist of the base pairs  
mostly contributes to the decrease of the geometry relaxation of each
nucleobase and in a reduction of the $\lambda_{i}$. The decrease of the 
energies of the adiabatic IP (see Fig. ~\ref{fig:fig5}) and the 
inner-sphere reorganization energy (see Fig. ~\ref{fig:fig2}) provide
the decrease of the vIP, which is the sum of above two components.

As we mentioned above, $\lambda_{i}$ depends on the charge distribution 
over the chain. The electrostatic potential distribution in the (G-C)$_n$ 
and (A-T)$_n$ oligomers and respectively the residence of the HOMO in 
the oligomer centers provides the localization of the charge on the 
central guanines and adenines in the oxidized state. We have calculated 
the charge distribution as the difference between oxidized $E_{+(A,D)}$ 
and neutral states $E_{(A,D)}$ with Mulliken population analysis 
\cite{Mulliken}. In the (G-C)$_n$ and (A-T)$_n$ sequences the charge is 
distributed along the chain and is characterized by the low charge density 
at the DNA molecule sides. For example, the density of the atomic partial 
charge localized on the $n$=1 site is lower than that at the chain center 
by 0.06 coul for the (A-T)$_n$ and by 0.25 coul for the (G-C)$_n$ sequences.  
For the (G-C)$_4$ sequence our results are in agreement with the data 
in Ref. \cite{alex}.

\begin{figure}
\includegraphics[scale=0.9]{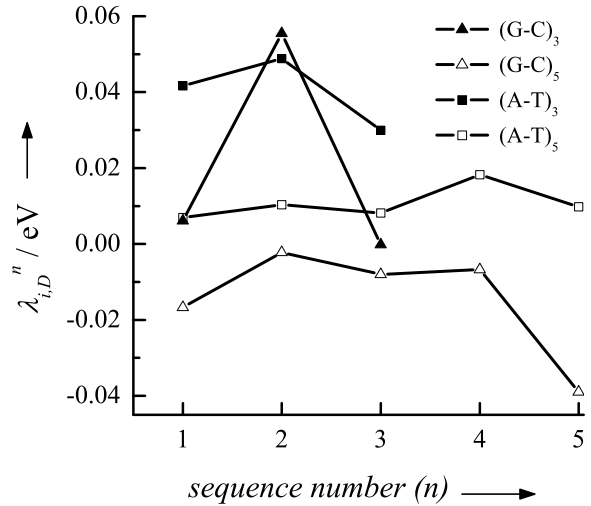}
\caption{The inner-sphere reorganization energy $\lambda_{i,D}$ 
corresponding to the single base pairs within the (A-T)$_n$ and 
the (G-C)$_n$ oligomers, where $n=3$ and $n$=5 are calculated with 
UB3P86/6-311$^{+}$G$^{\ast}$.}
\label{fig:fig3} 
\end{figure}

Therefore, the charge accumulation in the oligomer centers in the 
oxidized state produces the maximum geometry relaxation in the 
center of the DNA chain. We have performed an estimation of the geometry 
relaxation of the separated base pairs $\lambda_{i,D}^{n}$ within the 
optimized geometries of the (A-T)$_n$ and (G-C)$_n$ oligomers, where 
$n=1\ldots 6$. The simulation results of $\lambda_{i,D}^{n}$ for $n$=3 
and $n$=5 are presented in Figure ~\ref{fig:fig3}. Clearly, for the 
(G-C)$_n$ sequences the difference of the structure relaxation at the 
sides of the chain and in the center is significant than that for the 
(A-T)$_n$ sequences. The behavior of these curves repeats primarily 
the charge distribution in the (A-T)$_n$ and (G-C)$_n$ sequences.

The difference between the inner-sphere reorganization energy of 
the (A-T)$_n$ and (G-C)$_n$ oligomers should provides the larger 
magnitude of the vibrational coupling constant for the G-C pairs 
than that for the A-T pairs, and larger for the guanine than that 
for the adenine (see Table~\ref{tab:table1}).

\section{Conclusions} 
We have performed accurate quantum-chemical calculations to determine 
the electron coupling and the inner-sphere reorganization energy for the 
(A-T)$_n$ and (G-C)$_n$ DNA oligomers, where $n=1\ldots 6$. The electronic 
coupling between the two neighbor nucleobases within the same strand 
decreases exponentially with increasing of the base pairs number $n$ 
participating in the chain formation. The $n\ge$4 is the sequence number 
required for an accurate evaluation of the electron coupling in the DNA 
molecule. The orbital distribution in oligomers with the HOBO 
residing on the central nucleobase have been found to be the main reason 
for charge accumulation on the base pair located close to the chain center. 
The charge distribution in the chain determines degree of the the geometry 
relaxation of the base pair during the oxidation process in dependence on 
their location within the oligomer. Therefore, the base pairs in the chain 
center have stronger geometry distortion during the oxidation process. 
Such results are in good agreement with the theory of polaron formation in the 
DNA molecule, where the maximum structure distortion occurs in the polaron center 
\cite{bishop}. 

\section{Acknowledgment}
The authors would like to thank  Dr. E.B. Starikov for useful discussion.
This work has been supported by the Canada Research
Chair Program and a Canadian Foundation for Innovation (CFI) grant.

\end{document}